\documentclass{article}

\usepackage[english]{babel}

\usepackage[letterpaper,top=2cm,bottom=2cm,left=3cm,right=3cm,marginparwidth=1.75cm]{geometry}

\usepackage{graphicx}

\title{\textbf{A novel multi-threaded web crawling model}}
\author{Author: Weijie Jiang \\ Institution: Shanghai Maritime University}

\begin{document}
\maketitle

\begin{abstract}
This paper proposes a novel model for web crawling suitable for large-scale web data acquisition. This model first divides web data into several sub-data, with each sub-data corresponding to a thread task. In each thread task, web crawling tasks are concurrently executed, and the crawled data are stored in a buffer queue, awaiting further parsing. The parsing process is also divided into several threads. By establishing the model and continuously conducting crawler tests, it is found that this model is significantly optimized compared to single-threaded approaches.
\end{abstract}

\section{INTRODUCTION}

With the rapid development of the Internet, the volume of online information is experiencing explosive growth. Consequently, the demand for effective acquisition and management of online information is becoming increasingly urgent.

Web crawlers can automatically retrieve various types of information from the Internet. For individual users, web crawlers help users quickly access and integrate massive information resources. For enterprises, the role of web crawlers enables them to obtain product information from competitors, allowing for more effective market strategies. For governments, web crawlers assist in detecting trends in online public opinion and maintaining cybersecurity. Meanwhile, academic researchers can also use web crawlers to crawl academic resources and data, which aids in the advancement of scientific research.

Therefore, web crawlers, as an automated data acquisition tool, are becoming increasingly important in this context.

\section{MOTIVATION}
Traditional single-threaded web crawlers may encounter efficiency issues in large-scale data retrieval tasks due to their performance limitations. Therefore, to enhance the performance and efficiency of web crawlers, introducing multi-threading technology is a common solution. Multi-threaded web crawlers allow simultaneous processing of multiple tasks, with each task executed in an independent thread, thus fully utilizing computational resources and accelerating data retrieval speed.

Although multi-threaded web crawler technology is relatively mature today, often we only need to extract specific data from web pages after crawling. Based on this, this paper aims to develop a multi-threaded web crawler and a web page information extraction model, enabling users to retrieve large amounts of truly needed information in a short period.

\section{MODEL ARCHITECTURE}
\begin{figure}[htbp]
\centering
\includegraphics[width=0.8\linewidth]{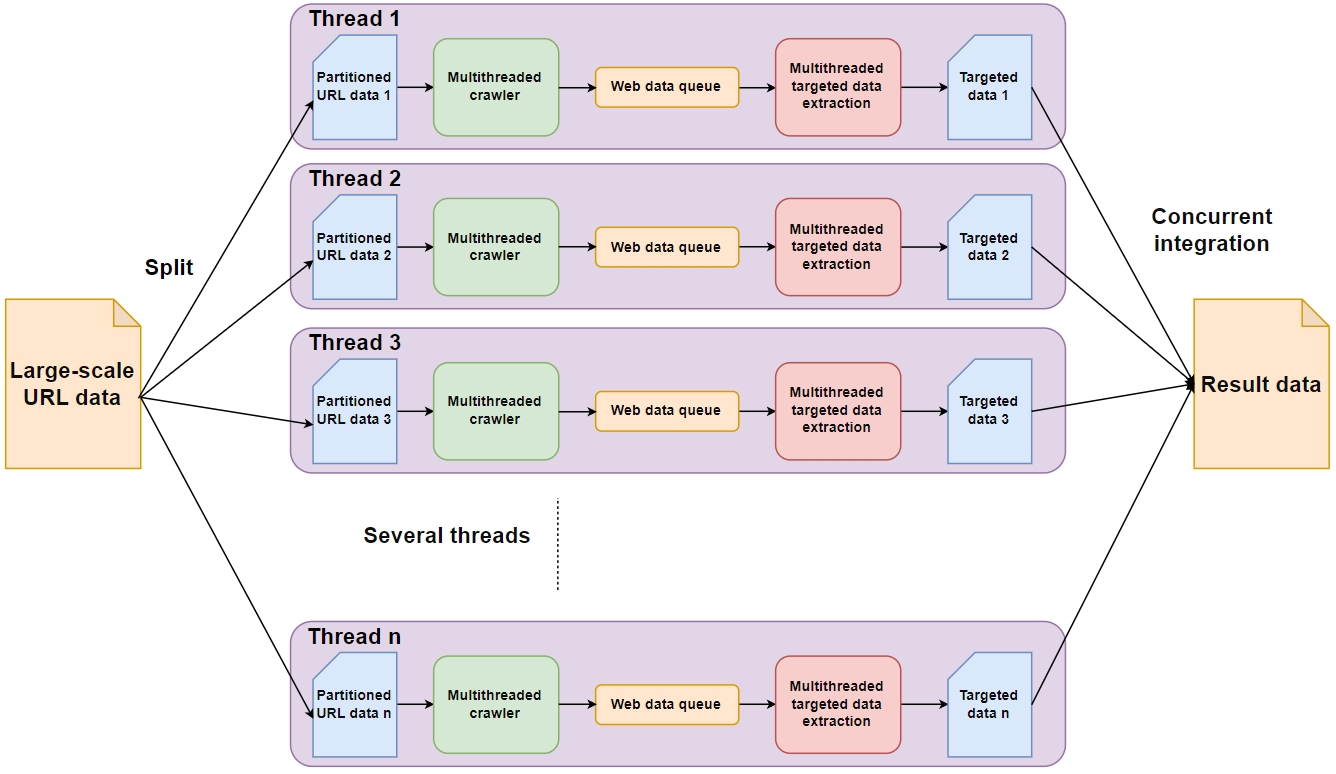}
\caption{\label{fig:model}Model Structure.}
\end{figure}

Figure \ref{fig:model} illustrates the model I constructed. Initially, the large text data is divided into several segments, each processed by a dedicated thread. Within these threads, URL data from the large text is concurrently crawled. The crawling process remains concurrent throughout. Subsequently, the crawled webpage text is stored in a web data queue, awaiting targeted data extraction. The targeted data extraction process also operates concurrently. Finally, the extracted data is written into targeted data sets. Each targeted data set is ultimately integrated into a result data set and returned. The writing process is concurrent as well.

\section{EXPERIMENT}
Firstly, we test the efficiency of the crawler under single-threaded conditions. Here, we crawl URL datasets with data sizes of 100, 500, and 1000 respectively, recording the time taken for each single-threaded execution. Each task size experiment is repeated ten times (Table \ref{tab:single_thread_results}).
\begin{table}[htbp]
\centering
\begin{tabular}{cccc}
\hline
Experiment & Data Size 100 & Data Size 500 & Data Size 1000 \\
\hline
1 & 17.562s & 92.442s & 247.962s \\
2 & 15.809s & 91.730s & 252.434s \\
3 & 18.474s & 94.995s & 248.887s \\
4 & 15.934s & 100.061s & 251.991s \\
5 & 16.210s & 95.860s & 250.444s \\
6 & 16.759s & 93.467s & 255.241s \\
7 & 15.475s & 91.966s & 252.567s \\
8 & 16.006s & 104.441s & 246.241s \\
9 & 16.273s & 95.511s & 255.124s \\
10 & 16.567s & 93.426s & 256.342s \\
Average Time & 16.507s & 95.391s & 251.724s \\
\hline
\end{tabular}
\caption{Crawler Time Cost under Single-threaded Conditions}
\label{tab:single_thread_results}
\end{table}

\begin{figure}[htbp]
\centering
\includegraphics[width=0.8\linewidth]{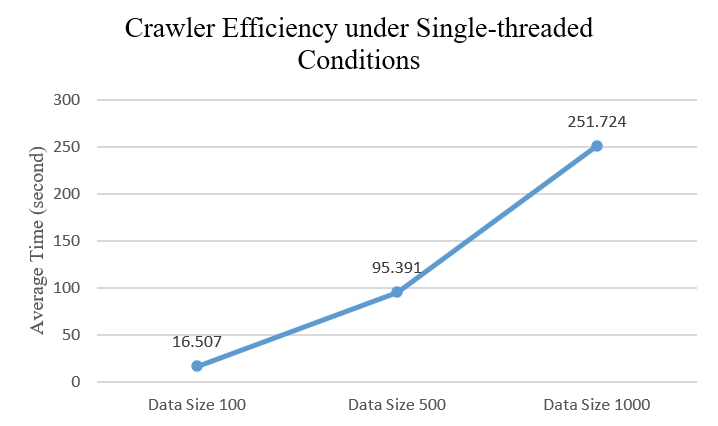}
\caption{\label{fig:single-threaded}Crawler Efficiency under Single-threaded Conditions.}
\end{figure}

Here we plot a line graph of data size against average time (Figure \ref{fig:single-threaded}), showing an overall linear relationship between data size and average time. It can be observed that as the data size increases, the time taken by the single-threaded crawler program exceeds linear growth. One possible reason for this is that with larger data sizes, the impact of network fluctuations becomes more significant.

Next, we proceed with executing multi-threaded crawling operations. Here, we maintain a constant number of channels in each thread, with 
\(m\)=5 and \(k\)=5. To simplify the experiment, we only test the scenario with a data size of 500. We conduct multiple experiments (10 in each group) to obtain the average values for text segmentation \(n\) of 1, 5, 10, and 20 (Table \ref{tab:thread_counts}).
\begin{table}[htbp]
\centering
\begin{tabular}{cccccc}
\hline
Experiment & Thread Count 1 & Thread Count 5 & Thread Count 10 & Thread Count 20\\
\hline
1 & 23.589s & 21.307s & 19.189s & 18.891s \\
2 & 24.145s & 22.013s & 19.616s & 18.762s \\
3 & 23.988s & 21.921s & 19.652s & 19.306s \\
4 & 23.892s & 21.340s & 18.965s & 19.441s \\
5 & 23.784s & 22.552s & 18.798s & 19.351s \\
6 & 24.185s & 22.498s & 19.987s & 20.534s \\
7 & 24.812s & 21.093s & 19.445s & 19.675s \\
8 & 25.009s & 22.062s & 19.213s & 18.321s \\
9 & 23.512s & 21.402s & 18.872s & 20.001s \\
10 & 23.902s & 22.512s & 19.765s & 19.842s \\
Average Time & 24.082s & 21.870s & 19.350s & 19.412s \\
\hline
\end{tabular}
\caption{Average Time for Different Thread Counts When \(n=10\), \(m=k=5\)}
\label{tab:thread_counts}
\end{table}

We observe that when \(m\)=5 and \(k\)=5, the shortest execution time is achieved when the thread count is set to 10, followed by 20. Due to the relatively small difference in time between thread counts 10 and 20, we speculate that the system's execution time stabilizes when the thread count is 10 or greater.

Table \ref{tab:thread_time} records the start, end, and duration of each thread in a concurrent experiment (\(n=10\), \(m=5\), \(k=5\)). The timestamp is May 8, 2024, at 18:55 Beijing time. In this experiment, there are 10 threads with 5 crawler channels and 5 data parsing channels.
\begin{table}[htbp]
\centering
\begin{tabular}{cccc}
\hline
Thread ID & Start Time & End Time & Duration \\
\hline
T0 & 18:55:03.7s & 18:55:23.1s & 19.358s \\
T1 & 18:55:03.7s & 18:55:23.2s & 19.419s \\
T2 & 18:55:03.7s & 18:55:23.0s & 19.238s \\
T3 & 18:55:03.8s & 18:55:23.2s & 19.395s \\
T4 & 18:55:03.8s & 18:55:23.1s & 19.339s \\
T5 & 18:55:03.8s & 18:55:23.2s & 19.363s \\
T6 & 18:55:03.8s & 18:55:23.1s & 19.3s \\
T7 & 18:55:04.0s & 18:55:23.0s & 18.965s \\
T8 & 18:55:04.0s & 18:55:22.7s & 18.611s \\
T9 & 18:55:04.1s & 18:55:23.0s & 18.886s \\
\hline
\end{tabular}
\caption{Thread Execution Time}
\label{tab:thread_time}
\end{table}

\begin{figure}[htbp]
\centering
\includegraphics[width=0.8\linewidth]{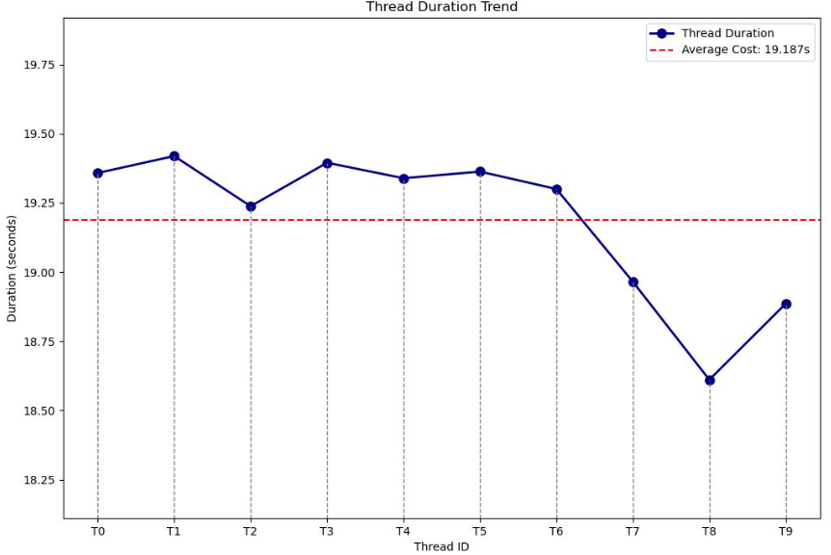}
\caption{\label{fig:n10m5k5}Thread Duration When \(n=10\), \(m=5\), \(k=5\)}
\end{figure}

The line graph in Figure \ref{fig:n10m5k5} visualizes the duration of each thread as recorded in Table \ref{tab:thread_time}.
Next, we change the value of \(m\) and \(k\) to 10 and proceed with the same experiment.
When \(n=10\), the time remains the smallest among the others. With the increase in the number of crawler and parsing channels, the overall program execution time reduces. However, it is worth noting that when \(n=20\), the time unexpectedly increases. Analyzing the reasons behind this, it is found that it may be due to the significant overhead of thread creation and switching, which reduces overall performance (Table \ref{tab:thread_counts}).
\begin{table}[htbp]
\centering
\begin{tabular}{cccccc}
\hline
Experiment & Thread Count 1 & Thread Count 5 & Thread Count 10 & Thread Count 20\\
\hline
1 & 19.092s & 18.676s & 18.543s & 23.346s \\
2 & 19.324s & 18.523s & 17.752s & 23.607s \\
3 & 18.984s & 17.982s & 17.635s & 20.087s \\
4 & 20.102s & 18.002s & 18.032s & 19.652s \\
5 & 19.767s & 17.613s & 17.424s & 20.142s \\
6 & 19.324s & 17.982s & 18.432s & 19.542s \\
7 & 19.645s & 18.654s & 17.733s & 21.421s \\
8 & 19.426s & 18.768s & 18.032s & 20.442s \\
9 & 19.816s & 18.982s & 17.998s & 22.421s \\
10 & 20.042s & 19.112s & 18.692s & 19.884s \\
Average Time & 19.552s & 18.429s & 18.027s & 21.054s \\
\hline
\end{tabular}
\caption{Average Time for Different Thread Counts When \(n=10\), \(m=k=10\)}
\label{tab:thread_counts2}
\end{table}
\section{SUMMARY}
After multiple experiments, it was found that when the number of URLs is 500, and 
\(n=m=k=10\), the least time is consumed, indicating the best scenario, with a time of 18.027 seconds. Comparing this to the single-threaded crawler time for the same dataset, the time is reduced by 77.364 seconds, optimizing by 81.11\%.
\begin{figure}[htbp]
\centering
\includegraphics[width=0.8\linewidth]{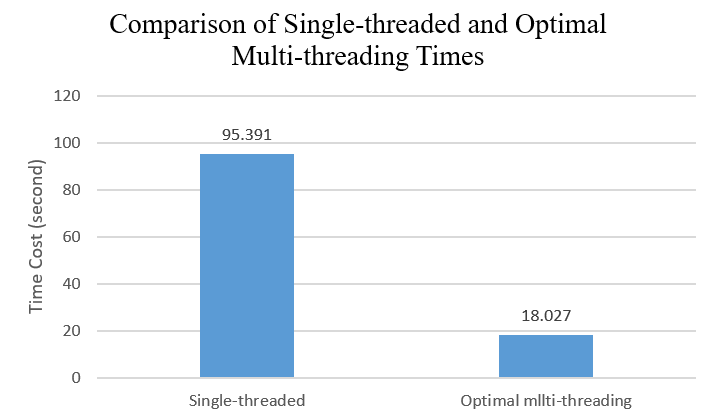}
\caption{\label{fig:comparsion}Comparison of Single-threaded and Optimal Multi-threading Times}
\end{figure}
\section{FUTURE WORK}
Obviously, this experiment is not yet fully completed. The preliminary conclusions drawn only provide guidance for our subsequent experiments. Next, we need to conduct controlled variable experiments multiple times, obtaining datasets with different numbers of URLs, and varying the values of 
\(n\), \(m\), and \(k\). 
By continuously adjusting the thread numbers in the model, we aim to determine the optimal thread settings applicable to all URL dataset sizes.
\newpage
\nocite{*}
	\bibliography{references} 
	\bibliographystyle{IEEEtran} 
\end{document}